\documentclass[final]{svjour2}
\usepackage{graphicx}
\usepackage{rotating}
\usepackage{amssymb}
\usepackage{mathptmx}
\usepackage[numbers]{natbib}
\makeatletter \journalname{Journal of Low Temperature Physics}

\begin{document}

\newcommand{\hdblarrow}{H\makebox[0.9ex][l]{$\downdownarrows$}-}
\title{Model calculation of orientational effect of deformed aerogel on the order parameter of superfluid $^3He$}

\author{E.V. Surovtsev \and I.A. Fomin}

\institute{P.L.Kapitza Institute for Physical Problems\\ 119334, Moscow, Russia \\
Tel.: +7(495)-1373248\\ Fax: +7(495)-6512125\\
\email{fomin@kapitza.ras.ru} }
\date{25.07.2007}

\maketitle


\begin{abstract}
Theory of Rainer and Vuorio of small objects in superfluid $^3He$
is applied for calculation of the average orientational effect of
a deformed aerogel on the order parameter of $^3He$. The minimum
deformation which stabilizes the ordered state is evaluated both
for specular and diffusive scattering of quasiparticles by the
threads of aerogel.

\end{abstract}

\section{Introduction}
Influence of high porosity aerogel on superfluid phases of $^3He$
phenomenologically can be described by the extra term in the free
energy functional:
\begin{equation}
\label{GL} F_{GL}=N(0)\int d^3 r \left[\eta_{jl}(\textbf{r})A_{\mu
j}A^{*}_{\mu l}+...\right],
\end{equation}
where $N(0)$ is the density of states on Fermi level, $A_{\mu j}$
is the order parameter of the superfluid $^3He$,
$\eta_{ij}(\textbf{r})$ is a real symmetric random tensor; it
describes a local anisotropy. Small changes of physical properties
of superfluid $^3He$ brought about by aerogel can be expressed in
terms of the ensemble average of the random tensor
$<\eta_{ij}(\textbf{r})>$ and of a correlation function
$P_{jlmn}(\textbf{r})=<\eta_{jl}(0)\eta_{mn}(\textbf{r})>$
$-<\eta_{jl}(0)><\eta_{mn}(\textbf{r}>$. The isotropic part of the
average tensor $<\eta_{jl}>^I\sim\delta_{jl}$ determines the
suppression of temperature of superfluid transition by aerogel.
And its anisotropic part
$<\eta_{jl}>^a=<\eta_{jl}>-1/3\eta_{nn}<\delta_{jl}>$ describes
orientational effect of a global anisotropy on the order
parameter. In case of the ABM-order parameter $A_{\mu
j}=(\Delta/\sqrt{2})d_\mu(m_j+in_j)$ orientational energy
$f_l=-N(0)(\Delta^2/2)$ $<\eta_{ij}>^al_il_j$, where
$\textbf{l}=\textbf{m}\times \textbf{n}$. For evaluation of the
mentioned averages aerogel can be envisioned as an array of
randomly oriented threads of a diameter $d\simeq 3\div5$ nm. This
diameter fits in the interval $a \ll d \ll\xi_0$, where $a$ is the
interatomic distance and $\xi_0=\hbar v_F/2\pi k_B T_{c}$ is the
correlation length for superfluid $^3He$. This condition allows to
consider threads of aerogel as "small objects" and apply the
theory of Rainer and Vuorio \cite{RV1}. Volovik \cite{Volovik}
used  results of this theory and a simple model of aerogel for
order of magnitude estimation of the effect of deformation on a
state of the A-like phase. This estimation is very sensitive to a
choice of parameters characterizing the system. In an attempt to
make the estimation less ambiguous using the same model we made
here explicit calculations of $<\eta_{jl}(\textbf{r})>$ in case of
a small uniaxial deformation of aerogel. We found also the
long-wavelength limit of Fourier-image of the correlation function
for a non-deformed aerogel.

\section{Model and Calculations}
We assume that aerogel consists of straight pieces of thread with
the length of $\varepsilon$ homogeneously distributed over the
volume. Threads have circular section with radius $\rho$.
Orientational distribution of threads is described by the
distribution function $n(\textbf{m})$, which is equal to the
number of threads in a unit volume having direction $\textbf{m}$.
For a non-deformed aerogel this function is isotropic:
\begin{equation}
n(\textbf{m})=\frac{1}{2\pi^2}\frac{1-P}{\rho^2\varepsilon},
\end{equation}
where $P$ is aerogel porosity. Uniaxial deformation of aerogel can
be described as a transformation of coordinates $z\rightarrow
(1+\gamma)z$, $x\rightarrow (1-\gamma/2)x$, $y\rightarrow
(1-\gamma/2)y$. Where $\gamma=\Delta L/L$ is deformation in
$z$-direction and positive sign of $\gamma$ corresponds to the
stretching along $z$-direction. After this transformation function
$n(\textbf{m})$ takes the form:
\begin{equation}
\label{n}
n(\textbf{m})=\frac{(1-P)}{2\pi^2\rho^2\varepsilon}\cdot\left(1+\frac{\gamma}{2}(3
\cos^2(\theta)-1)\right),
\end{equation}
where $\theta$ is the angle between $\textbf{z}$-axis and
direction $\textbf{m}$.

According to the Rainer and Vuorio theory a small object gives
rise to additional term in the free energy:
\begin{equation}
\label{FE} \Delta F^{Obj}=\int d^3 r ~d^3 r^{'}~ A^{*}_{\mu
j}(\textbf{r})K^{Obj}_{jl}(\textbf{r},\textbf{r}^{'})A_{\mu
l}(\textbf{r}^{'}),
\end{equation}
The kernel $K^{Obj}_{jl}$ has a form:
\begin{eqnarray}
\label{Kernel} K^{Obj}_{jl}(\textbf{r},\textbf{r}^{'})=\frac{2\pi
N(0)k_B T_c}{\hbar v_F}\sum_n
\exp\left(-2|\omega_n|\frac{|\textbf{r}_0-\textbf{r}|+|\textbf{r}_0-\textbf{r}^{'}|}{v_F}\right)\times\nonumber\\
\times
\frac{1}{4\pi}\frac{(\textbf{r}_0-\textbf{r})_j(\textbf{r}_0-\textbf{r}^{'})_l}{|\textbf{r}_0-\textbf{r}|^3|\textbf{r}_0-\textbf{r}^{'}|^3}
\left[\frac{d\sigma}{d\Omega}(\textbf{r}_0\hat{-}\textbf{r},\textbf{r}_0\hat{-}\textbf{r}^{'})-\delta(\textbf{r}_0\hat{-}
\textbf{r},\textbf{r}_0\hat{-}\textbf{r}^{'})\sigma_{tot}(\textbf{r}_0\hat{-}\textbf{r})\right],
\end{eqnarray}
where $\omega_n=(2n+1)\pi k_B T/\hbar$, $\textbf{r}_0$ is
coordinate of the object, $d\sigma/d\Omega$, $\sigma_{tot}$ --
differential and total cross-section of the scattering
quasiparticles by the object, respectively. Knowing $K^{Obj}_{jl}$
one can obtain $\eta_{jl}$ as \cite{F1}:
\begin{equation}
\label{etta} \eta_{jl}(\textbf{r})=\frac{1}{N(0)}\int d^3
r^{'}K^{Obj}_{jl}(\textbf{r},\textbf{r}^{'}).
\end{equation}
For application of  Eq.(\ref{Kernel}) to the piece of thread with
the length $\varepsilon\sim\xi_0$ it should be divided on elements
with the length $\delta\varepsilon\ll\xi_0$. If one introduces
average distance between the threads $\xi_a$ then perturbation of
the order parameter produced by a single element at a distance of
the order of $\xi_a$ is small as $\rho\delta \varepsilon/\xi_a^2$.
The result for the whole piece of thread in a principal order on
$\rho/\xi_a$ can be found as a sum of contributions from all
elements provided that this sum remains small correction to the
non-perturbed order parameter. Condition of smallness can be
violated in a vicinity of a thread,  but contribution of this
region to the integral (\ref{etta}) is of the order of
$\rho^2/\xi_0^2$ and therefore does not introduce considerable
error. Contribution to $\eta_{jl}(\textbf{r})$ from all threads in
a principal order on their concentration is found by summation of
contributions from each piece of thread. Cross-sections
$d\sigma/d\Omega$ and $\sigma_{tot}$ entering Eq.(\ref{Kernel})
depend on character of quasiparticle scattering by the threads. We
consider diffusive and specular scattering as two extreme cases.
\begin{figure}
\begin{center}
\includegraphics[%
  width=0.75\linewidth,
  keepaspectratio]{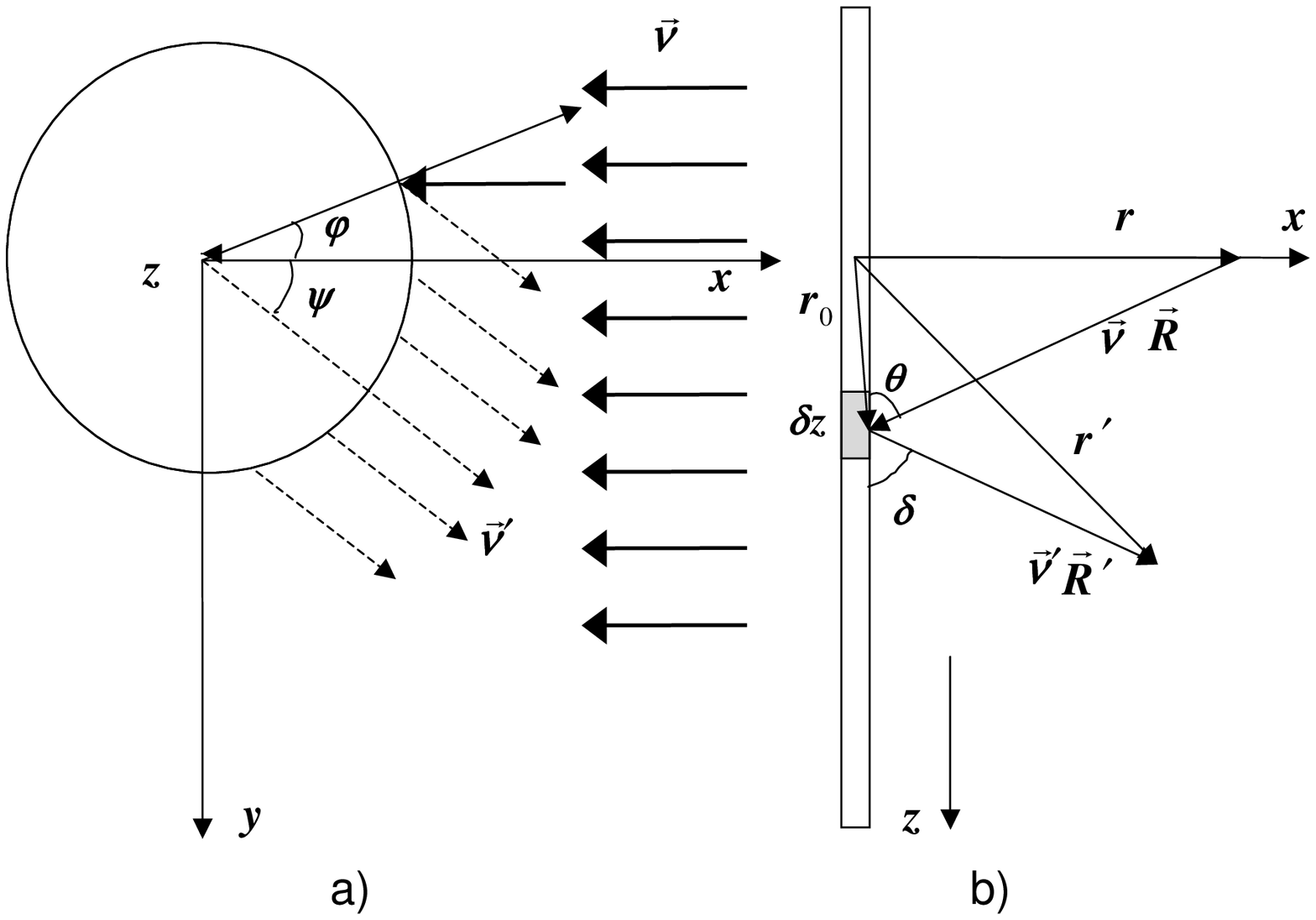} 
\end{center}
\caption{a) - quasiparticle scattering by the thread, view from
above, b) - a frame of references, side view } \label{frame}
\end{figure}

Let us introduce a frame of references as shown in a
Fig.(\ref{frame}). In this frame differential cross-section of a
quasiparticle by an element of the thread $\delta z$ at the
diffusive scattering has the form:
\begin{equation}
\frac{d\sigma}{d\Omega}(\nu,\nu^{'})=\rho\cdot \delta
z\cdot\sin(\theta) \cdot \frac{1+cos(\psi)}{2\pi},
\end{equation}
where
$\nu=\frac{\textbf{r}_0-\textbf{r}}{|\textbf{r}_0-\textbf{r}|}$ ,
$\nu^{'}=\frac{\textbf{r}^{'}-\textbf{r}_0}{|\textbf{r}_0-\textbf{r}^{'}|}$.
Then for the total cross-section we have: $
\sigma_{tot}=2\rho\sin(\theta)\delta z.$ Substitution of these
expressions into Eqns.(\ref{Kernel}, \ref{etta}) and integration
over $\delta z$ yields the following expression for the
contribution from a single thread $\eta^{1}_{jl}(\textbf{r})$:
\begin{equation}
\eta^1_{jl}(\textbf{D},\textbf{m})=d_jd_l \eta_{x}^1(D)+m_jm_l
\eta_{z}^1(D),
\end{equation}
where $\textbf{D}$ is the perpendicular from the point of
observation to the thread or its continuation, $\textbf{m}$ and
$\textbf{d}$ - unit vectors along the thread and along
$\textbf{D}$ respectively. Functions $\eta^{1}_{x}(\textbf{r})$
and $\eta^{1}_{z}(\textbf{r})$ are given by the following
expressions:
\begin{eqnarray}
\eta^{1}_{x}(\frac{x}{2\xi_0}\textbf{r})=-\int_{h/2\xi_0-\varepsilon/4\xi_0}^{h/2\xi_0+\varepsilon/4\xi_0}
\frac{\rho}{\xi_0}\cdot\frac{\alpha^2}{(\zeta^2+\alpha^2)^2}
\ln\left(\tanh(\zeta^2+\alpha^2)^{1/2}\right)
\cdot\nonumber \\
\cdot\left[\frac{1}{16}+\frac{\alpha}{2\pi(\zeta^2+\alpha^2)^{1/2}}
\right]d\zeta,\\
\eta^{1}_{z}(\textbf{r})=-\int_{h/2\xi_0-\varepsilon/4\xi_0}^{h/2\xi_0+\varepsilon/4\xi_0}
\frac{1}{2\pi} \frac{\rho}{\xi_0}\cdot\frac{\alpha
\zeta^2}{(\zeta^2+\alpha^2)^{5/2}}\ln\left(\tanh(\zeta^2+\alpha^2)^{1/2}\right)d\zeta,
\end{eqnarray}
where $\alpha=D/(2\xi_0)$, $h$ is a distance from the center of
the thread to the plane of observation. Contribution from all
threads is:
\begin{equation}
\eta_{jl}(\textbf{r})=\sum_s
\eta_{jl}^{s}(\textbf{D}^s,\textbf{m}^{s}),
\end{equation}
where $s$ is a number of the thread. As a result of averaging one
arrives at the isotropic part of tensor $\eta_{jl}$:
\begin{equation}
\label{iso}
<\eta_{jl}>^{I}=\frac{8}{3}(X+Z)\frac{(1-P)\xi_0}{\rho}\delta_{jl}.
\end{equation}
Anisotropic part for a uniaxial deformation can be written as:
\begin{eqnarray}
{\kappa}\equiv\label{ani}- <\eta_{xx}>^a=
-<\eta_{yy}>^a=1/2<\eta_{zz}>^a
=\frac{4}{15}(2Z-X)\frac{(1-P)\xi_0}{\rho}\gamma.
\end{eqnarray}
Coefficients in Eqns.(\ref{iso},\ref{ani}) are expressed in terms
of the integrals:
\begin{equation}
X=\int_{0}^{\infty}\int_0^{\infty}
\eta_{x}~\frac{dh}{\varepsilon}\alpha
d\alpha,~Z=\int_{0}^{\infty}\int_0^{\infty}
\eta_{z}~\frac{dh}{\varepsilon} \alpha d\alpha.
\end{equation}
Numerical evaluation of these integrals shows that at
$\varepsilon\geq0.002\xi_0$ the answer does not practically depend
on $\varepsilon$. For $\varepsilon\sim\xi_0$ one can take values
of $X$ and $Z$ obtained in a limit $\varepsilon\rightarrow\infty$,
when they can be evaluated analytically $X^D=13\pi^2/768$,
$Z^D=\pi^2/256$. Then
\begin{equation}
\label{ani1}
<\eta_{jl}>^{I}=\frac{\pi^2}{18}\frac{(1-P)\xi_0}{\rho}\delta_{jl},\qquad
{\kappa^D}=-\frac{7}{20}\left(\frac{\pi}{12}\right)^2\frac{(1-P)\xi_0}{\rho}\gamma.
\end{equation}
Corrections to the order parameter brought up by fluctuations are
expressed in terms of  the Fourier-image of correlation function
at $\textbf{k}\rightarrow0$, i.e. $\int P_{jlmn}(\textbf{r})d^3r$
\cite{F2}. From symmetry considerations it can be written as:
\begin{equation}
\int
P_{jlmn}(\textbf{r})d^3r=\Phi_0(\delta_{jm}\delta_{ln}+\delta_{jn}\delta_{lm}-\frac{2}{3}\delta_{jl}\delta_{mn}).
\end{equation}
The calculations analogous to that made when finding $<\eta_{jl}>$
yield the following expression for $\Phi_0$:
\begin{equation}
\Phi_0=\frac{32\pi}{15}\varepsilon\xi_0^2(X-2Z)^2(1-P).
\end{equation}
For diffusive scattering, using for $X^{D}$ and $Z^{D}$ their
values for infinite threads we have:
\begin{equation}
\label{Phi0D}
\Phi_0^D=\frac{49}{270}\left(\frac{\pi}{4}\right)^5\varepsilon\xi_0^2(1-P)\approx
0.054\varepsilon\xi_0^2(1-P).
\end{equation}
For specular boundary conditions differential cross-section has
the form:
\begin{equation}
\frac{d\sigma}{d\Omega}=\frac{\rho \delta
z}{2}\cos(\frac{\psi}{2})\sin(\theta)\delta(\nu_z{'}-\nu_z).
\end{equation}
Repeating with this cross-sections all calculations one arrives at
$<\eta_{jl}>^I$ and ${\kappa^S}$ ($X^S=\pi^2/64$):
\begin{equation}
\label{aniS}
<\eta_{jl}>^I=\frac{\pi^2}{24}\frac{(1-P)\xi_0}{\rho}\delta_{jl},\qquad
{\kappa^S}=-\frac{\pi^2}{270}\frac{(1-P)\xi_0}{\rho}\gamma,
\end{equation}
i.e. in comparison with diffusive scattering anisotropic part of
tensor $<\eta_{jl}>$ is approximately at $1.7$ times greater. For
correlation function in the case of specular scattering the answer
is:
\begin{equation}
\label{Phi0S}
\Phi_0^S=\frac{8}{15}\left(\frac{\pi}{4}\right)^5\varepsilon\xi_0^2(1-P)\approx
0.159\varepsilon\xi_0^2(1-P).
\end{equation}
\section{Discussion}
The considered model gives for orientational energy of the order
parameter of the ABM-phase in a uniaxially deformed aerogel
\begin{equation}
E_l^D=\frac{7}{120}\left(
\frac{\pi}{4}\right)^2N(0)\Delta^2\frac{(1-P)\xi_0}{\rho}\gamma
l_z^2,
\end{equation}
in the case of diffusive scattering and
\begin{equation}
E_l^S= \frac{\pi^2}{180}N(0)\Delta^2\frac{(1-P)\xi_0}{\rho}\gamma
l_z^2,
\end{equation}
for specular reflection. Effect of anisotropy is greater for the
specular reflection, but in both cases the energy $E_l$ is one-two
orders of magnitude smaller than the simple order of magnitude
estimation cf.\cite{Volovik}. Even greater difference occurs for
the borderline deformation $\gamma_c$ separating the state of the
ordered ABM-phase from the state with the critical fluctuations.
Transition between the two states is expected when the
fluctuational corrections to the equation, determining a form of
the order parameter become comparable with $\Delta^2$. According
to \cite{F2} that happens when
$\frac{5\sqrt{5}}{32\pi}\Phi_0/(\xi_0^3\sqrt{{\kappa}})\simeq1$.
Using for $\Phi_0$ and $\kappa$ expressions (\ref{ani1}),
(\ref{Phi0D}) we have for the diffusive scattering:
\begin{equation}
\gamma_c^D=\frac{25}{12}(2Z^D-X^D)^3\frac{\rho}{\xi_0}\frac{\varepsilon^2}{\xi_0^2}(1-P)\simeq
-1.5\cdot10^{-3}\frac{\rho}{\xi_0}\frac{\varepsilon^2}{\xi_0^2}(1-P)
\end{equation}
If we assume that $\varepsilon=\xi_a$ and substitute for $\rho=2$
nm and $\xi_a=16$ nm, then $\gamma_c^D\simeq-1.2\cdot10^{-5}$.
With the expressions (\ref{aniS}), (\ref{Phi0S}) and the same
values of parameters for the specular reflection:
\begin{equation}
\gamma_c^S=\frac{25}{12}(2Z^S-X^S)^3\frac{\rho}{\xi_0}\frac{\varepsilon^2}{\xi_0^2}(1-P)\simeq
-7.4\cdot10^{-3}\frac{\rho}{\xi_0}\frac{\varepsilon^2}{\xi_0^2}(1-P)\simeq
-6\cdot10^{-5}
\end{equation}
 Specular scattering renders higher limit for
$\gamma_c$, but even in this case it is about two orders of
magnitude smaller than the simple estimation \cite{Volovik}.
Present calculations are based on a simple model, so we can not
claim that the obtained results describe effect of the aerogel on
superfluid $^3He$ quantitatively. Nevertheless the observed
suppression of the effect of anisotropy is hardly an artifact of
the model.

\begin{acknowledgements}
This research was supported in part by RFBR grant (no.
07-02-00-214), by Ministry of Education and Science of Russian
Federation, Russian Science-Support Fund, Landau Scholarship
(A.F.) from Forschungszentrum J\"{u}lih, Germany and CRDF (grant
RUP1-2632-MO04)
\end{acknowledgements}

\end{document}